\documentclass[osajnl,preprint,showpacs,superscriptaddress]{revtex4}
\usepackage{bm,graphicx,amsmath}
\usepackage{pstricks}

\newcommand{\bra}[1]{\ensuremath{\langle#1|}}
\newcommand{\Ket}[1]{\ensuremath{|#1\rangle}}
\newcommand{\ket}[1]{\ensuremath{|#1\rangle}}

\begin{document}

\title{Cavity-QED entangled photon source based on\\
two truncated Rabi oscillations}
\author{R.~Garc\'{i}a-Maraver, K.~Eckert, R.~Corbal\'{a}n, and J.~Mompart}
\affiliation{Departament de F\'{i}sica, Universitat Aut\`{o}noma de Barcelona, E-08193 Bellaterra, Spain}
\date{\today}

\begin{abstract}
We discuss a cavity-QED scheme to deterministically generate entangled photons pairs by using a three-level atom successively coupled to two single longitudinal mode high-Q cavities presenting polarization degeneracy. The first cavity is prepared in a well defined Fock state with two photons with opposite circular polarizations while the second cavity remains in the vacuum state. A half-of-a-resonant Rabi oscillation in each cavity transfers one photon from the first to the second cavity, leaving the photons entangled in their polarization degree of freedom. The feasibility of this implementation and some practical considerations are discussed for both, microwave and optical regimes. In particular, Monte Carlo wave function simulations have been performed with state-of-the-art parameter values to evaluate the success probability of the cavity-QED source in producing entangled photon pairs as well as its entanglement capability.
\end{abstract}

\pacs{42.50.Pq, 03.67.Mn, 32.80.-t}

\maketitle

\section{INTRODUCTION}
Entanglement is a quantum correlation without classical counterpart that
appears in composite quantum systems and
constitutes one of the main resources in quantum communication and
quantum information processing. In particular, entangled photon pairs
have been considered for testing quantum mechanics
against local hidden variable theories \cite{LHV} as well as for quantum information
applications such as teleportation \cite{teleportation}, dense coding
\cite{coding}, and quantum key distribution (QKD) \cite{QKD}.
In fact, the main issue in cryptography is the secure distribution
of the encoding key between two partners. For this purpose,
quantum cryptography renders two classes of protocols
\cite{BB84,SARG04,Ek91}
based, respectively, on superposition and quantum measurement, and
entanglement and
quantum measurement that, under certain requirements, are proven to be
unconditionally secure by physical laws. Thus, superposition based
protocols
require the use of a deterministic single photon source to be
unquestionably
safe under attacks of a third party. In fact, the first implementation of
the BB84 protocol \cite{BB84}
using a true single photon source was reported only recently \cite{Gr00}.

On the other hand, entanglement based protocols were first considered by A.~Ekert
\cite{Ek91} and have been experimentally implemented by means of parametric down converted photons generated in non-linear
crystals \cite{QCE1,QCE2,QCE3}.
In all these cases the statistics of the photon number and time distributions follows,
essentially, a poissonian law. Therefore, in order to reduce the
number of multiphoton pairs, the average photon number has to be much less than
one which, in turns, strongly reduces the key exchange rate. Hence, one of the
practical issues in entanglement based quantum cryptography presently attracting
more attention \cite{ZSS05,YYG05} is the development of light sources that emit
deterministically single entangled photon pairs at a constant rate.
In this context, we will discuss in this paper a cavity quantum electrodynamics (cavity-QED) 
proposal for the deterministic generation of 
polarization entangled photon pairs.  

Cavity-QED with Rydberg atoms crossing
superconducting microwave resonators \cite{Har,Wal} or a single atom/ion
strongly coupled to a high-Q optical cavity \cite{Rem,Kug,Kim,Oro}
provide close to ideal physical systems for quantum state
engineering of both, the atom's electronic degrees of freedom and the intracavity field
\cite{Hei,Wal,Mor05}. We propose here a cavity-quantum electrodynamics
(cavity-QED) implementation \cite{Har,Wal,Kim,Oro,optfock,Hei} that, making
use of a three-level atom coupled successively to two high-Q cavities, allows
for the deterministic generation of polarization entangled photon pairs.
The initial separable state of the system composed of the atom and the
two degenerate modes of each cavity is choosen such that the
relevant couplings of the system can be reduced to those of a
three-level interaction between the initial state and a bright state
\cite{Ari} involving the two excited atomic states and the modes of 
first cavity, and thereafter between this bright state and the polarization entangled
photon state. Thus, the entangling mechanism consists in the
implementation of two spatially separated $\pi$-Rabi oscillations with
the two polarization modes of each cavity such that the atom transfers
one photon from the first to the second cavity while entangling their
polarization degree of freedom.

In Section II we present the physical model under investigation while
the entangling mechanism is sketched in Section III. In Section IV we introduce the Hamiltonian of the system and discuss its coherent dynamics. Effects of decoherence are taken into account by means of the Monte Carlo wave function approach in Section V. Several figures of merit describing the entanglement capability of the cavity-QED source are introduced and evaluated in Section VI. In Section VII we briefly discuss the
main physical requirements of our implementation. Finally, we
present the conclusions of this article in Section VIII.

\section{PHYSICAL FRAMEWORK}

The physical system under study is sketched in Fig.~1. It consists of a three-level atom or ion with its two electric dipole allowed transition frequencies denoted by $\omega _{ac}$ and $\omega _{bc}$ and two high-Q cavities labeled $1$ and $2$, both supporting a single longitudinal mode of equal angular frequency $\omega _{c}$ and displaying polarization degeneracy, i.e., we deal with one three-level atom and four cavity modes for the e.m. field. The optical axes of the cavities will be taken as the quantization axis for the angular momentum.

\begin{figure}[h]
	\centering
		\includegraphics[width=0.39\textwidth]{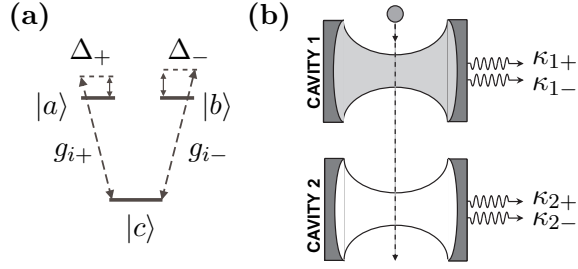}
		\begin{picture}(0,0)(0,0)
\put(-131,45){$ g_{i-}$}
\put(-182,45){$ g_{i+}$}
\put(-155,15){$\ket{c}$}
\put(-176,82){$\Delta_+$}
\put(-139,82){$\Delta_-$}
\put(-189,61){$\ket{a}$}
\put(-126,61){$\ket{b}$}
\put(-1,80){$\kappa_{1+}$}
\put(-1,70){$\kappa_{1-}$}
\put(-1,27){$\kappa_{2+}$}
\put(-1,19){$\kappa_{2-}$}
\put(-199,+95){{\bf (a)}}
\put(-109,+95){{\bf (b)}}
\end{picture}
\caption{Setup for the generation of polarization entangled photon pairs. It consists of a three-level atom or ion with electric dipole transitions frequencies $\omega_{ac}$ and $\omega_{bc}$ coupling, respectively, to the $\sigma_{+}$ and $\sigma_{-}$ circular polarizations of each of the two high-Q cavities. Both cavities support the same single longitudinal mode frequency $\omega_c$. $\kappa_{i\pm}$ with $i=1,2$ is the photon decay rate through the mirrors for the $\sigma_\pm$ circular polarization of the corresponding cavity.}
	\label{fig:micro2}
\end{figure}

We assume that the atomic transitions $\ket{c}\leftrightarrow\ket{a}$ and $\ket{c}\leftrightarrow\ket{b}$, satisfying the electric dipole selection rules, are coupled to the longitudinal optical mode of each cavity via the two opposite circular polarizations $\sigma _{\pm }$, e.g., a $J=0$ to $J=1$ transition.
The strength of the couplings is given by \cite{libro}
\begin{eqnarray}
\Omega_{i \pm }(t)=\sqrt{ (2 g_{i\pm}(t) \sqrt{n_{i\pm} + 1} )^2 + \Delta_{\pm}^2 },
\end{eqnarray}
where $i=1,2$ indexes the cavity, and $\pm$ the polarization. $g_{i\pm}(t)$ is the vacuum Rabi frequency of the respective cavity mode, $n_{i\pm}$ is the number of photons in the corresponding cavity mode, and $\Delta _{+}\equiv\omega _{c}-\omega _{ac}$\ and $\Delta _{-}\equiv\omega _{c}-\omega _{bc}$ are the detunings. We will consider next the completely symmetric case given by $\Delta _{+}=\Delta _{-} (\equiv\Delta)$ and
$g_{i+}=g_{i-} (\equiv {g_i})$ and, eventually, we will take into account experimental imperfections that relax the former symmetry conditions.

\section{ENTANGLING MECHANISM}
We assume the ability to prepare the intracavity fields in a Fock state \cite{Wal} and take 
$\ket{\psi(t=0)}=a^{\dag}_{1+} a^{\dag}_{1-} \ket{\Omega} \equiv \ket{I}$ as the initial state of the system with 
$\ket{\Omega} \equiv \ket{c}\otimes \ket{\Omega}_1 \otimes \ket{\Omega}_2 $. $\ket{\Omega}_i$ is the two mode vacuum state of cavity $i$, with $a^{\dag}_{i\pm}$ ($a_{i\pm}$) being the photon creation (annihilation) operator in the corresponding cavity mode. We denote the two atomic lowering operators as $S_+ = \ket{c}\bra{a}$ and $S_- = \ket{c}\bra{b}$. Then, to generate a single entangled photon pair, we will assume \emph{first} that the system is initially prepared into the separable state $\ket{I}$, i.e., the atom is in the ground state and the e.m. field is in a Fock state with cavity $1$ confining one $\sigma_+$-circularly polarized photon and one $\sigma_-$-photon, and cavity $2$ containing no photons.
\begin{figure*}[h]
	\centering
		\includegraphics[width=0.63\textwidth]{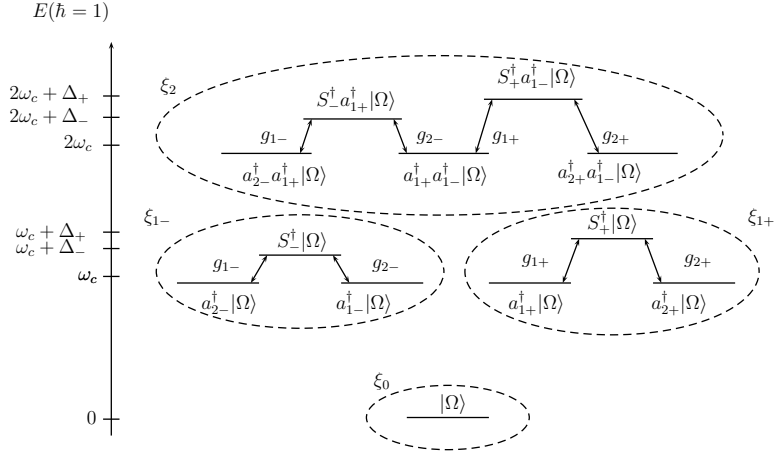}
	\label{fig:fig2}
			\caption{States of the system formed of the three-level atom and the two modes of each cavity grouped into manifolds (ellipses) according to Hamiltonian (\ref{H}-\ref{H2}). Only the lowest energy manifolds are shown. In the absence of incoherent processes, the continuous coherent evolution in each manifold is decoupled from the rest. $g_{i}(t)$ is the quantum Rabi frequency of cavity $i$ and $\ket{\Omega} \equiv \ket{c}\otimes \ket{\Omega}_1 \otimes \ket{\Omega}_2 $, being $\ket{\Omega}_i$ the two mode vacuum state of cavity $i=1,2$.}
	\label{fig:energies}
\end{figure*}
\emph{Second}, the atom interacts with the e.m.~field of cavity $1$ where it can likewise absorb the $\sigma_{+}$ or the $\sigma_{-}$ photon. This dynamics can be easily understood in terms of the bright-dark states \cite{Ari}. In fact, under the two-photon resonance condition, i.e., $\Delta_{+}=\Delta_{-}$, state $\left| c\right\rangle $ couples only to the particular combination of states $\sqrt{2} \ket{B} \equiv  \left( S^{\dag}_+ a^{\dag}_{1-} + S^{\dag}_- a^{\dag}_{1+} \right) \ket{\Omega}$, the bright state, while remaining uncoupled, due to destructive interference, to the dark state, $\sqrt{2} \ket{D} \equiv  \left( S^{\dag}_+ a^{\dag}_{1-} - S^{\dag}_- a^{\dag}_{1+} \right) \ket{\Omega}$. Therefore, this three-level system behaves effectively as a two-level system with Rabi oscillations occurring between  states $\left|I\right\rangle$ and $\left| B\right\rangle$. With this picture in mind, the strength of the coupling and/or the interaction time in cavity $1$ must be adjusted to yield half-of-a-resonant Rabi oscillation such that, after this process, the population is completely transferred from $\ket{I}$ to $\ket{B}$. Except for a global phase, the final state of the system after this step is $\sqrt{2} \ket{B}= \left( S^{\dag}_+ a^{\dag}_{1-} + S^{\dag}_- a^{\dag}_{1+} \right) \ket{\Omega}$.

\emph{Third}, the atom interacts with the vacuum modes of cavity $2$. The strength and interaction time is taken now such that the excited atom yields a photon in this cavity, i.e., again half of a 
Rabi oscillation is performed. In fact, the two pathways $ S^{\dag}_+ a^{\dag}_{1-}\ket{\Omega} \stackrel{}{\rightarrow}a^{\dag}_{2+}a^{\dag}_{1-} \ket{\Omega}$ and $ S^{\dag}_- a^{\dag}_{1+}\ket{\Omega} \stackrel{}{\rightarrow}a^{\dag}_{2-}a^{\dag}_{1+} \ket{\Omega}$ add constructively to yield the following final state: $\sqrt{2} \ket{E^+}\equiv  \left( a^{\dag}_{2+}a^{\dag}_{1-}  + a^{\dag}_{2-}a^{\dag}_{1+}  \right) \ket{\Omega}$. Hence, at the end of the overall process, if the $\sigma_{-}$ photon remains in cavity $1$ then the $\sigma_{+}$ photon has been transferred to cavity $2$ and vice versa. Therefore, the polarization modes of the first cavity become entangled with those of the second cavity. 

Note that by associating one qubit to each cavity such that state $\ket{0}$ of the qubit corresponds to having a $\sigma_+$ photon and state $\ket{1}$ to having a $\sigma_-$ photon in the cavity, the final state of the cavity field corresponds to the two-qubit Bell state $\ket{\psi^+}=\frac{1}{\sqrt{2}}\left( \ket{1}\otimes\ket{0} + \ket{0}\otimes\ket{1}\right)$.

\section{DYNAMICS OF THE SYSTEM}
In what follows we will analyze in detail the dynamics of the entanglement mechanism. To check the validity of the previous proposal we integrate the Schr\"odinger equation of the full system composed of the three-level atom and the four e.m.~modes. With this aim, we write down the Hamiltonian of the full system in the rotating wave approximation ($\hbar =1$), 
\begin{eqnarray}
	H_{T}=H_{0}+H_{I},\label{H}
\end{eqnarray}
\begin{eqnarray}
H_{0}=\sum_{i=1,2}\omega_{c} \left(  a_{i+}^{\dag}a_{i+}
+ a_{i-}^{\dag}a_{i-}\right)
+ \sum_{j=a,b} \omega_{jc}\ket{j}\bra{j}\label{H1}
\end{eqnarray}
\begin{eqnarray}
H_{I}=\sum_{i=1,2} g_{i}
\left( a_{i+}^{\dag} S_{+} + a_{i+} S_{+}^{\dag} + a_{i-}^{\dag} S_{-} +a_{i-} S_{-}^{\dag}\right)\label{H2}
\end{eqnarray}
Fig.~\ref{fig:energies} shows the energies and couplings arising from Eqs. (\ref{H}-\ref{H2}). The states coupled by the Hamiltonian have been grouped into manifolds $\xi_{0},\xi_{1-},\xi_{1+},\xi_{2}....$, labelled by the excitations, such that each manifold is decoupled from the rest. In the absence of decoherence, if the system starts in a certain state of a given manifold it will remain there during all its coherent evolution. Therefore, by taking $ \ket{I} = a^{\dag}_{1+} a^{\dag}_{1-} \ket{\Omega}$ as the initial state of the system, we can restrict ourselves to investigate the evolution of the system in manifold $\xi_{2}$. In the interaction picture, the Hamiltonian of the system restricted to $\xi_2$ takes the form:
\begin{eqnarray}
H&=&g_{1}e^{-i\Delta_{+} t}S^{\dag}_+ a^{\dag}_{1-}\left|\Omega\right\rangle \left\langle I\right|+
g_{1}e^{-i\Delta_{-} t}S^{\dag}_- a^{\dag}_{1+}\left|\Omega\right\rangle\left\langle I\right|+\nonumber \\
&&g_{2}e^{-i\Delta_{+} t}S^{\dag}_+ a^{\dag}_{1-}\ket{\Omega}\bra{\Omega} a_{2-}a_{1+} +
g_{2}e^{-i\Delta_{-} t}S^{\dag}_- a^{\dag}_{1+}\ket{\Omega} \bra{\Omega} a_{2+}a_{1-}+ h.c.\label{ham0} 
\end{eqnarray}
Let us consider now the alternative basis of $\xi_{2}$ given by states:
\begin{eqnarray}
\ket{I}&\equiv&a^{\dag}_{1+} a^{\dag}_{1-} \ket{\Omega}, \label{ba0}\\
\sqrt{2} \ket{B}& \equiv & \left( S^{\dag}_+ a^{\dag}_{1-} + S^{\dag}_- a^{\dag}_{1+} \right) \ket{\Omega}, \\
\sqrt{2} \ket{D}& \equiv & \left( S^{\dag}_+ a^{\dag}_{1-} - S^{\dag}_- a^{\dag}_{1+} \right) \ket{\Omega}, \\
\sqrt{2} \ket{E^+}& \equiv & \left( a^{\dag}_{2+}a^{\dag}_{1-}  + a^{\dag}_{2-}a^{\dag}_{1+}  \right) \ket{\Omega}, \\
\sqrt{2} \ket{E^-}& \equiv & \left( a^{\dag}_{2+}a^{\dag}_{1-}  - a^{\dag}_{2-}a^{\dag}_{1+}  \right) \ket{\Omega}.\label{basis}
\end{eqnarray}
Notice that in both states $\ket{E^{+}}$ and $\ket{E^{-}}$ the atom factorizes and the cavity fields are entangled according to the $\ket{\psi^{+}}$ and $\ket{\psi^{-}}$ Bell states, respectively. Using Eqs. (\ref{ba0}-\ref{basis}), Hamiltonian (\ref{ham0}) can be rewritten as:
\begin{eqnarray}
H&=& \left( \sqrt{2}g_{1}\ket{B}\bra{I} +g_{2}\ket{B}\bra{E^+} + g_{2}\ket{D}\bra{E^-} \right)
\times\cos \left( \left(\Delta_- - \Delta_+\right) t/2 \right)\nonumber\\
&&+ \left(\sqrt{2}g_{1}\ket{D}\bra{I}+g_{2}\ket{B}\bra{E^-} + g_{2}\ket{D}\bra{E^+} \right)\times\sin\left( \left(\Delta_- - \Delta_+\right) t/2 \right) +h.c.\label{hamil}
\end{eqnarray}
\begin{figure}[h]
\centering
\includegraphics[width=0.40\textwidth]{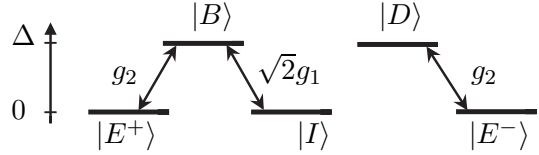}
\begin{picture}(0,0)(0,0)
\put(-95,-7){$\ket{I} $}
\put(-171,-7){$\ket{E^+} $}
\put(-30,-7){$\ket{E^-} $}
\put(-64,+38){$\ket{D} $}
\put(-135,+38){$\ket{B} $}
\put(-110,+17){$\sqrt{2}g_1$}
\put(-165,+17){$g_2$}
\put(-29,+17){$g_2$}
\put(-203,+2){$0$}
\put(-203,+30){$\Delta$}
\end{picture}
	\caption{The manifold $\xi_{2}$ of Fig. 2 expressed in terms of the states of the basis given in Eqs. (\ref{ba0}-\ref{basis}) and under the two-photon resonance condition.}
	\label{fig:esq}
\end{figure}

Therefore, under the two photon resonance condition, $\Delta_{+}=\Delta_{-}\equiv\Delta$, one finds that $\ket{D}$ is uncoupled from the initial state, i.e., $\bra{D}H\ket{I}=0$. Also $\bra{B}H\ket{E^{-}}=\bra{D}H\ket{E^{+}}=0$. The remaining couplings have been schematically represented in Fig.~(\ref{fig:esq}). In this case Hamiltonian (\ref{hamil}) simplifies to:
\begin{eqnarray}
H&=&\sqrt{2}g_{1}\ e^{-i\Delta t}\left| B\right\rangle\bra{I} + g_{2}\ e^{-i\Delta t}\ket{B}\bra{E^{+}}+ g_{2}\ e^{-i\Delta t}\ket{D}\bra{E^{-}}+ h.c.\label{Hn}
\end{eqnarray}
This suggests to shape $g_{1}(t)$ and $g_{2}(t)$ such that the system is transfered from $\ket{I}$ to $\ket{B}$ in the first cavity and from $\ket{B}$ to the entangled state $\ket{E^{+}}$ in the second cavity.  Explicitly, the dynamical evolution of the system will be obtained by integrating the Schr\"{o}dinger equation with Hamiltonian (\ref{Hn}). For the sake of making the discussion analytical, let us consider first that $g_i$ ($i=1,2$) is constant in time and different from zero only during a time interval of duration $t_i$. These two time intervals do not overlap and coupling occurs first in cavity 1. Therefore, in cavity $1$, the system evolves in time according to
\begin{eqnarray}
\ket{\psi(t)}&=&{e^{-i\Delta t/2}} \Big{[} \frac{-i2\sqrt{2}g_{1}}{\Omega _{B}} \sin(\Omega _{B}t/2)\Big{]}\ket{B}\nonumber\\
& & \hspace{-1.3cm}+ {e^{i\Delta t/2}} \Big{[}\cos(\Omega _{B}t/2) -i \frac{\Delta}{\Omega _{B}}\sin(\Omega _{B}t/2)\Big{]}\ket{I}\label{wf}
\end{eqnarray}
where $\Omega_{B}=\sqrt{8g_{1}^2+\Delta^2}$ is the generalized Rabi frequency. From Eq.(\ref{wf}) it is inferred that if the single photon resonance condition is fulfilled, i.e., $\Delta=0$, population is completely transferred from $\ket{I}$ to $\ket{B}$ if $\Omega_B t_1=\pi$, i.e., whenever half of a Rabi oscillation (a $\pi$-pulse) between these two states takes place. In this case, the quantum state will be:
\begin{eqnarray}
\hspace{-0.2cm}\ket{\psi(t_{1})}=-i\ket{B}=- \frac{i }{\sqrt{2}}\left( S^{\dag}_+ a^{\dag}_{1-} + S^{\dag}_- a^{\dag}_{1+} \right) \ket{\Omega}
\end{eqnarray}
After a time $t_{f}$ of free evolution, the atom interacts with the vacuum modes of cavity $2$ and the system evolves according to:
\begin{figure}[t]
	\centering
				\includegraphics[width=0.45\textwidth]{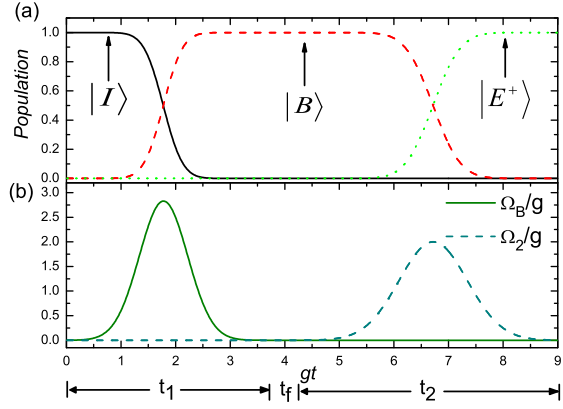}
	\caption{ (a) Coherent time evolution of the population of states:	$\ket{I}$ (solid curve), $\sqrt{2} \ket{B} \equiv  \left( S^{\dag}_+ a^{\dag}_{1-} + S^{\dag}_- a^{\dag}_{1+} \right) \ket{\Omega}$ (dashed curve), and $\sqrt{2} \ket{E^+} \equiv  \left( a^{\dag}_{2+}a^{\dag}_{1-}  + a^{\dag}_{2-}a^{\dag}_{1+}  \right) \ket{\Omega}$ (dotted curve) versus dimensionless time $gt$. (b) Time dependent generalized Rabi frequencies in dimensionless units for cavity $1$ (solid curve) and cavity $2$ (dashed curve). Parameters are: $\Delta_+=\Delta_-=0$, $gt_f=0.111$, $gt_{1}=1.110$ and $gt_{2}=1.570$. $g$ is the vacuum Rabi frequency at the center of each cavity mode. A gaussian temporal profile has been assumed for the coupling strength (see Eq.(\ref{gp})).}
\label{fig:evolu}
\end{figure}
\begin{eqnarray}
\ket{\psi(t)}&=&{e^{-i\Delta t'/2}} \Big{[}-\frac{2g_{2}}{\Omega _{2}}\sin(\Omega _{2}t'/2)\Big{]}\ket{E^{+}}\nonumber\\
& &\hspace{-1.3cm} -\frac{e^{i\Delta t'/2}}{2} \Big{[}  i\cos(\Omega _{2}t'/2)  +  \frac{\Delta }{\Omega _{2}}  \sin(\Omega _{2}t'/2)\Big{]}
\ket{B}\label{wff}
\end{eqnarray}
being $\Omega_{2}=\sqrt{4g_{2}^2+\Delta^2}$ the generalized Rabi frequency in the second cavity and $t'=t-(t_{1}+ t_{f})$. On single-photon resonance ($\Delta = 0$), if the interaction in the second cavity lasts a time $t_2$ fulfilling $\Omega_2 t_2=\pi$, then the population will be completely transferred from $\ket{B}$ to $\ket{E^{+}}$:
\begin{eqnarray}
\ket{\psi(t_{1}+t_f+t_{2})}&=&\frac{-1}{\sqrt{2}}\left( a^{\dag}_{2+}a^{\dag}_{1-}  + a^{\dag}_{2-}a^{\dag}_{1+}  \right) \ket{\Omega}\nonumber \\
&=&-\Ket{E^+} 
 \hspace{0.9cm}\label{es}
\end{eqnarray}
To be more realistic, we will consider next a gaussian profile for the atom-field interaction of the form:
\begin{eqnarray}
g_{i}(t)=	g e^{ -(t- \widetilde{t}_i)^2/\tau_i^2  }\quad {\rm with} \quad i=1,2, \label{gp}
\end{eqnarray}
where $\widetilde{t}_1=t_{1}/2$ and $\widetilde{t}_2=t_{2}/2 + t_f+t_1$. $t_i$ is the interaction time in each cavity, $t_f$ is the delay time (or free time of flight) between the two interactions and $\tau_i$ is the width of the corresponding gaussian profile. For the simulation shown in Fig.~\ref{fig:evolu} we have taken $\Delta=0$, $gt_1=1.11$, $gt_f=0.11$, $gt_2=1.57$, $g\tau_{1}=0.255$, and $g\tau_{2}=0.448$ being $g$ the quantum Rabi frequency at the center of each cavity that, for simplicity, we assume to be the same for all four cavity modes. Note that the previous parameters have been choosen such that $\int_0^{t_1}\Omega_B (t)dt=\pi$ and $\int_{t_1+t_f}^{t_1+t_f+t_2}\Omega_2 (t) dt=\pi$. Fig.~\ref{fig:evolu} (a) shows the numerical integration of the Schr\"odinger equation of the three-level atom successively interacting with the two cavities. One can clearly see from Fig.~\ref{fig:evolu}(a) that for $t=t_1$ the system is completely transfered to the bright state. Likewise, the second part of the process drawn at the right part of Fig.~\ref{fig:evolu}(a) shows that at the end of the complete process the two cavities become entangled in their polarization modes according to Eq.(\ref{es}).

\section{Monte Carlo Wave Function Simulations}
In the previous analysis only the coherent interaction of the atom with the modes of the cavities was considered. However, a realistic study of the feasibility of the present proposal has to take into account incoherent processes, i.e., dissipation and photon detection.
\begin{figure}[h]
	\centering
		\includegraphics[width=0.30\textwidth]{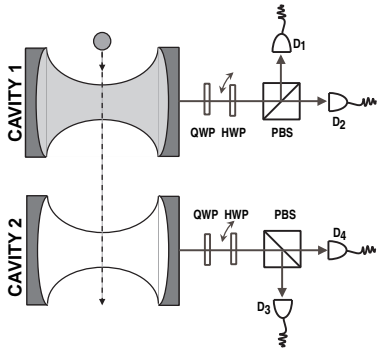}
\caption{Sketch of the basic optical elements needed for the Bell state analysis of the polarization entangled photon source. QWP: quarter wave plate; HWP: rotating half wave plate; PBS: polarization beam splitter, and D$_i$: single photon detector.}
\label{fig:micro2}
\end{figure}

We will discuss and characterize the cavity-QED source together with the detection system shown in Fig. 5. Let us assume that the quantum efficiency for the detectors is perfect ($\eta=1$). Two kinds of dissipative processes will be considered:
(i) Spontaneous atomic decay from the two optical transitions $\ket{a}$ to $\ket{c}$ and $\ket{b}$ to $\ket{c}$ at the common rate $\Gamma$ and (ii) cavity decay of the photons through the mirrors and the irreversible process of their detection. Since $\eta=1$, the parameters $\kappa_1\pm=\kappa_2\pm\equiv\kappa$ will denote the mirror transmission coefficients that, for simplicity, we take the same for all four cavity modes.

To account for these dissipative processes one could consider either the Liouville equation for the density operator of the system or the Monte Carlo wave function (MCWF) formalism.
In particular, the MCWF formalism is interesting at least for two reasons \cite{QJ1}: (i) in the MCWF treatment of a system belonging to a $N$ dimensional Hilbert space the number of real variables is $2N-1$ while the density matrix has $N^2-1$, and (ii) it provides new insights into the underlying physical mechanisms. In what follows we will use the MCWF formalism. 

The time evolution of the system, a so-called quantum trajectory, will be calculated by integrating the time-dependent Schr\"{o}dinger equation with the non-hermitian effective Hamiltonian $H_{\rm eff}$:
\begin{eqnarray}
H_{\rm eff}=H -i\sum_{i=1,2}\frac{\kappa}{2}(a^{\dag}_{i+}a_{i+} +a^{\dag}_{i-}a_{i-})  -i\frac{\Gamma}{2}\sum_{j=+,-}S_{j}^{\dag}S_{j}\hspace{1.5cm}
\end{eqnarray}
This Hamiltonian includes dissipation due to spontaneous decay of the atom at a rate $\Gamma$ as well as cavity decay at a rate $\kappa$.
\begin{figure*}[h]
	\centering
	\includegraphics[width=0.48\textwidth]{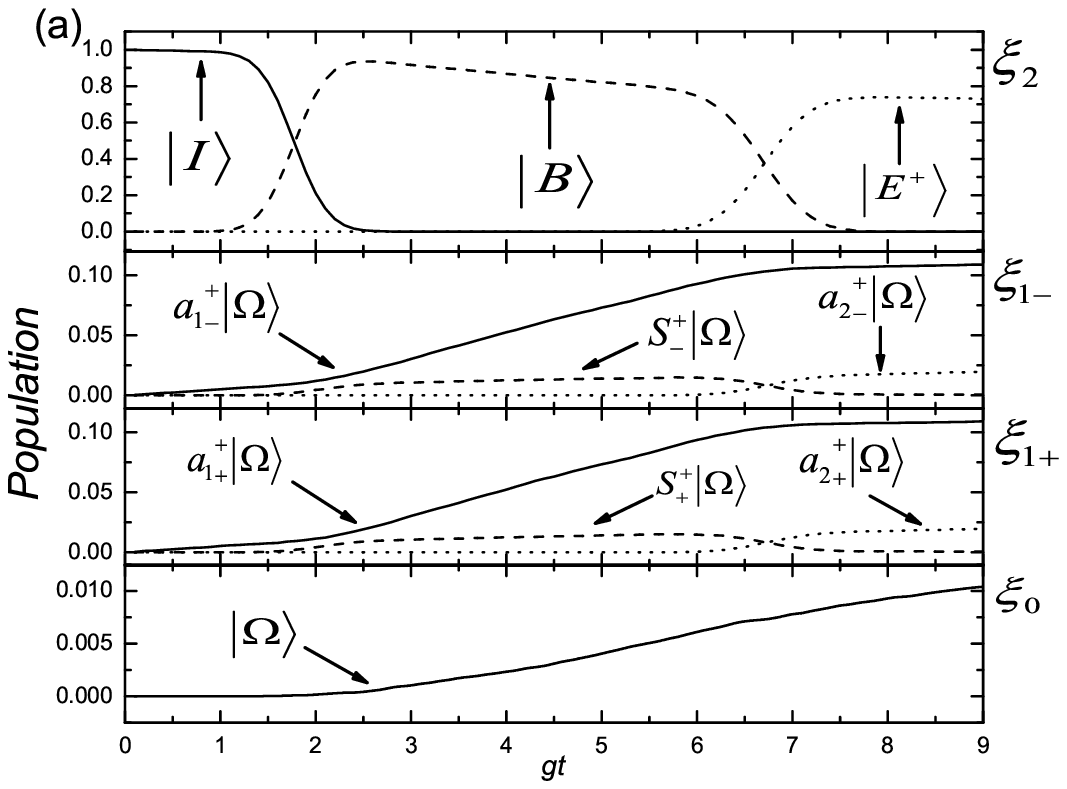}
	\includegraphics[width=0.47\textwidth]{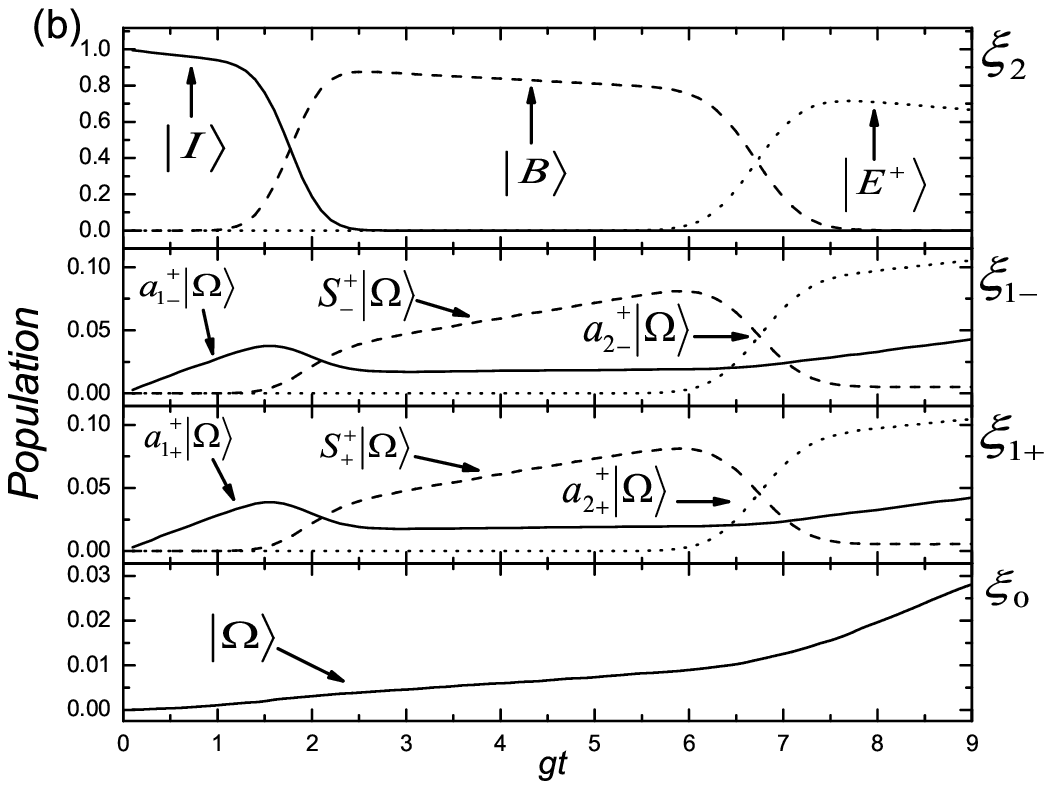}
	\caption{System evolution in presence of dissipative processes obtained averaging over many MCWF simulations with $\ket{\psi(0)}=\ket{I}$. Manifolds $\xi_{1-}$, $\xi_{1+}$ and $\xi_{0}$ become populated due to dissipative processes. The parameter setting is: (a) 	$\Gamma=0.05g$, $\kappa=0.1 \Gamma$, $\Delta=0$, and (b) $\kappa=0.03g$, $\Gamma=0.1 \kappa$, $\Delta=0$. The rest of parameters are the same as in Fig.~\ref{fig:evolu}.}
	\label{fig:evolMC1}
\end{figure*}

A quantum trajectory will consist of a series of continuous coherent evolution periods interrupted by quantum jumps occurring at random times. Averaging over many realizations of these quantum trajectories reproduces the ensemble results of the density matrix equations. The probability $dp$ that a quantum jump occurs in a time $dt$ is $dp=\sum_{n}G_{n} p _{n}dt$ where $G_{n}$ is the sum of the rates associated to all incoherent processes departing from state $n$ with population $p _{n}$. This summation is extended to all populated states. The time interval 
$dt \ll G_{n}^{-1}$ is choosen to assure that at most one quantum jump process occurs in this interval. At each interval $dt$ of time, a pseudo-random number $\epsilon\in [0,1]$ is used to determine whether a quantum jump takes place. For $\epsilon\geq dp$ no quantum jump occurs and the system evolves, after renormalization of the system wave function, according to the Schr\"{o}dinger equation with the non-Hermitian effective hamiltonian. In contrast when $\epsilon< dp$ a quantum jump takes place. If so, $\epsilon$ will be also used to decide which particular quantum jump process occurs proportionally to the rate of this process. The final state of the system will be fixed by the particular quantum jump that has taken place. Then, the continuous coherent evolution resumes again as well as the possibility of having more quantum jumps. 

Fig.~\ref{fig:evolMC1} shows the system evolution in the presence of dissipative processes obtained by averaging over many MCWF realizations. In Fig.~\ref{fig:evolMC1}(a), cavity decay is almost negligible and the dominant dissipative process is spontaneous atomic decay. Therefore, dissipation from manifold $\xi_2$ occurs only when the system is in the bright state $\ket{B}$ since in $\ket{I}$ and in $\ket{E^+}$ the atom is in its internal ground state. Accordingly, in the central region of Fig.~\ref{fig:evolMC1}(a) population is pumped into manifolds $\xi_{1-}$ and $\xi_{1+}$. Once in these manifolds a second dissipative process, mainly spontaneous emission (cavity decay of photons) at the center (end) of Fig.~\ref{fig:evolMC1}(a), brings population to the lowest energy manifold $\xi_0$. Clearly, to assure the maximum population of state $\ket{E^+}$ it is favourable to reduce $t_f$ as much as possible in order to minimize the overall time that the atom remains excited

On the other hand, in Fig.~\ref{fig:evolMC1}(b) the dominant dissipative process is the cavity decay of photons, being spontaneous atomic decay almost negligible. Thus, dissipation from manifold $\xi_2$ occurs during the whole process since all states of this manifold contain at least one photon. Since states $\ket{I}$ and $\ket{E^{+}}$ contain two photons while state $\ket{B}$ only one, the former decayes twice faster than the latter. In contrast, dissipation from manifolds $\xi_{1-}$ and $\xi_{1+}$ occurs mainly when the atom is in state $\ket{c}$, since those states in which the atom is excited have no photons. Consequently, population is significantly pumped out from $\xi_2$ at the beginning and at the end of the process.

\section{CHARACTERIZATION OF THE CAVITY-QED SOURCE}

\begin{figure}[b]
	\centering
		\includegraphics[width=0.40\textwidth]{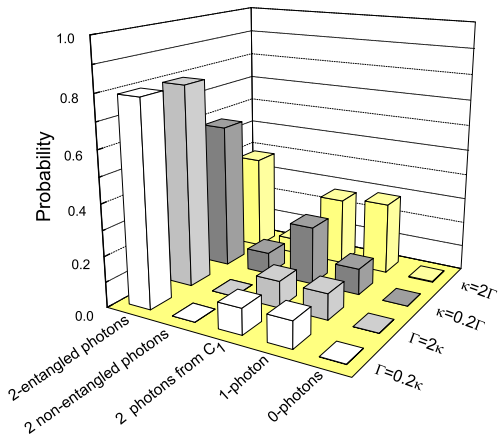}
\caption{ Probabilities for the different processes involving photon emission trough the cavity mirrors and their eventual photodetection. The sets of parameters correspond to $\kappa=0.2\Gamma, \kappa=2\Gamma$ for $\Gamma=0.05g$, $\Gamma=0.2\kappa$, $\Gamma=2\kappa$ for $\kappa=0.5g$, and $\Delta_+=\Delta_-=0$.}
	\label{fig:statistics}
\end{figure}

\begin{figure*}[t]
	\centering
		\includegraphics[width=0.32\textwidth]{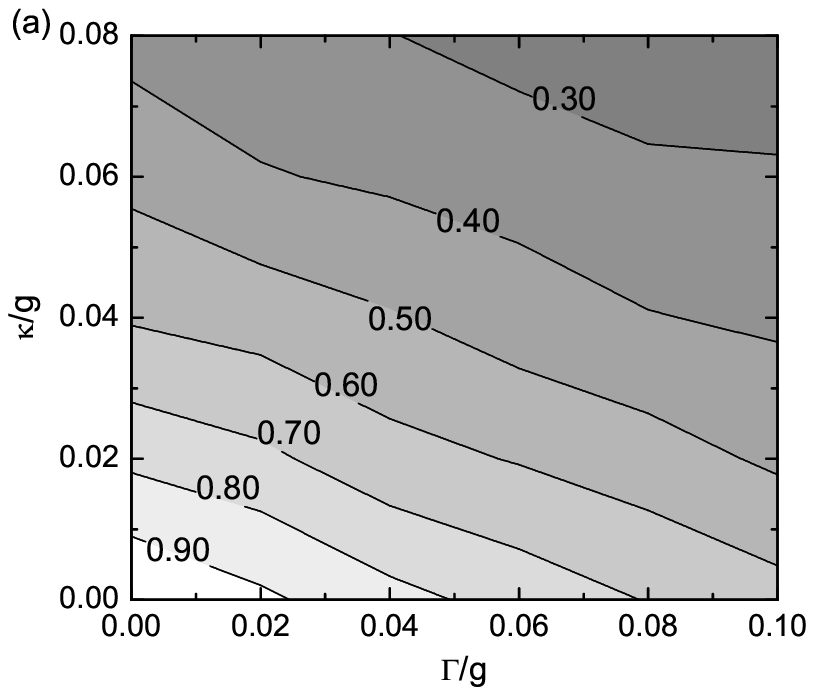}
		\includegraphics[width=0.32\textwidth]{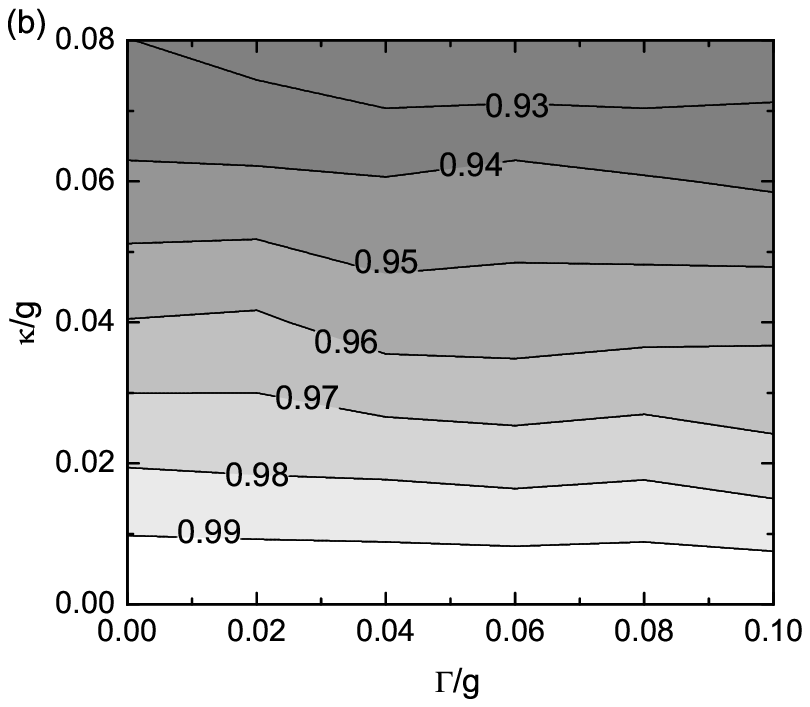}
		\includegraphics[width=0.32\textwidth]{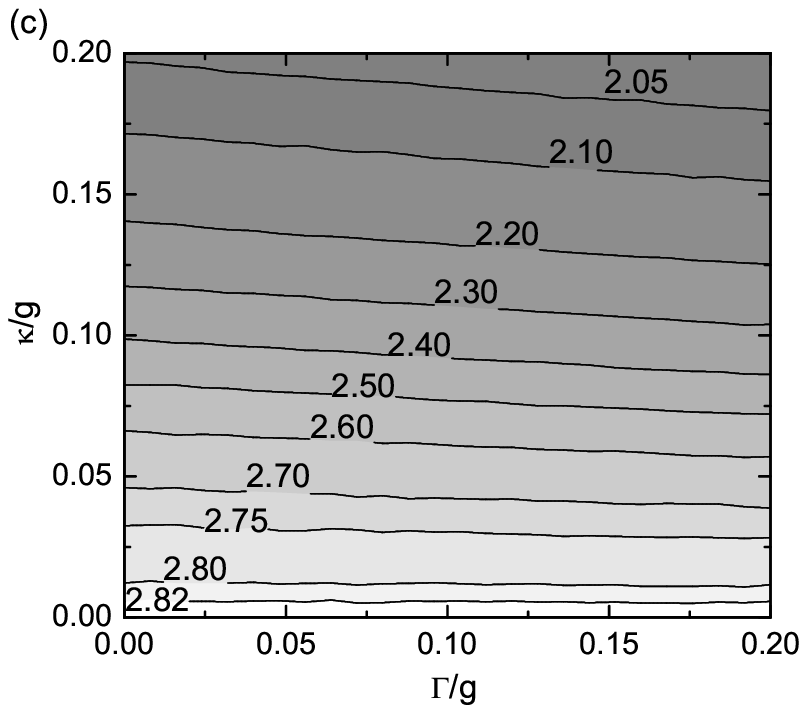}
\caption{Contour plots of (a) the success probability $P$, (b) the fidelity $F$ and (c) the $S$ parameter (c).
Parameters are the same as in Fig.(\ref{fig:evolu}).}
	\label{fig:f5}
\end{figure*}

Next we will characterize the cavity-QED source through three parameters: (i) the success probability $P$ of producing the state $\ket{E^{+}}$ after sending one atom through the setup, (ii) the fidelity $F$, and (iii) the $S$ parameter of the CHSH inequality \cite{CHSH} to quantify the entanglement capability of the source. Fidelity and $S$-parameter will be evaluated after post-selection of events which yield one photon from each cavity. Throughout the analysis we assume that the quantum efficiency for the detectors is perfect ($\eta=1$). MCWF simulations allows one to make a statistical analysis to obtain the probabilities of the different physical processes giving rise to (i) no cavity emitted photons, (ii) a single cavity emitted photon, two photons emitted from (iii) the same cavity or (iv) from different cavities but in a separable state, and (v) two entangled cavity emitted photons. In Fig.~(\ref{fig:statistics}), the probabilities for these processes to happen are shown for various  sets of parameters.
The leftmost column corresponds to the the success probability $P$ of producing an entangled pair of photons after sending one atom through the cavities. This value is calculated with respect to the full Hilbert space containing all the possible events (i-v). On the other hand, post-selection of events with one photon leaving each cavity allows to reduce to the space of two qubits defined by the polarizations of the photons. Let $p_{\rm 2 ph}$ be the probability for coincidence photodetection, {\it i.e.}, $p_{\rm 2 ph}$ is the sum of probabilities of events (iv) and (v). After post-selection we calculate the fidelity of the source as
\begin{eqnarray}
F=  \bra{E^{+}}\rho\ket{E^{+}} \label{F},
\end{eqnarray}
where 

\begin{eqnarray}
\rho=(1-\alpha)\frac12\left( a^{\dag}_{2+}a^{\dag}_{1-}\ket{\Omega}\bra{\Omega}a^{}_{2+}a^{}_{1-}
 + a^{\dag}_{2-}a^{\dag}_{1+}\ket{\Omega}\bra{\Omega}a^{}_{2-}a^{}_{1+}   \right )+
\alpha\ket{E^+}\bra{E^+}
\end{eqnarray}

is the final density matrix of the system and $\alpha=P/p_{\rm 2 ph}$. In addition, we use the $S$ parameter of the Bell-CHSH inequality \cite{CHSH} to characterize the entanglement capability. In particular, $S=2\sqrt{2}$ for maximally entangled states and $S=\sqrt{2}$ for a non-entangled state.\\
Fig.~\ref{fig:f5} shows the three figures of merit $P$, $F$, and $S$ as a function of cavity decay rate $\kappa$ and atomic decay rate $\Gamma$. Since photons can leak out the cavities at any time of the process, while spontaneous atomic decay can occur only when the atom is excited, the loss of photons becomes the dominant decoherence process. This is reflected in  Fig.~\ref{fig:f5} (a), where the success probability is smallest in the situation in which the loss of photons is the dominant dissipation mechanism. On the other hand, the fidelity and the $S$ parameter, calculated after reducing to the two-qubit space, present a different behavior as seen in Fig.~\ref{fig:f5} (b,c). The entangled state $\ket{E^{+}}$ as well as states giving rise to a non-entangled pair of photons, can decay only through the loss of photons since the atom is in the stable state $\ket{c}$. Thus $F$ as well as $S$ become almost insensitive to the spontaneous atomic decay.

\begin{figure}[h]
	\centering
			\includegraphics [width=0.33\textwidth]{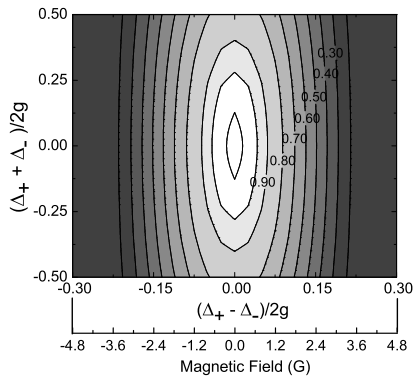}
	\caption{Success probability of the entangled photon pair source as a function of the deviation from the single and the two-photon resonance condition. The parameters are as in Fig.\ref{fig:evolu} with $g/2\pi=34$\rm{MHz}.}
	\label{fig:EB1}
\end{figure}

So far we have only discussed the completely resonant case $\Delta_+=\Delta_-\equiv0$. For certain experimental imperfections, this condition will no longer be fulfilled. In fact, the presence of a stray magnetic field will break the two-photon resonance condition. Assuming that $\ket{a}=\ket{J=1, m_{J}=+1}$, $\ket{b}=\ket{J=1, m_{J}=-1}$, and $\ket{c}=\ket{J=0, m_{J}=0}$, a magnetic field will break the degeneracy between atomic states $\ket{a}$ and $\ket{b}$, such that the two photon resonance condition is no longer satisfied, {\it i.e.}, $\Delta_+\neq \Delta_-$. The relation between the strength of a magnetic field and deviation from the two photon resonance condition is given by $\hbar \Delta_{\pm}=\pm \mu_{B} g_{J}B$, with $g_J$ being the gyromagnetic factor, $\mu_B$ being the Bohr magneton, and $g_J=3/2$ for the case chosen here. In this situation, it follows from from Eq. ({\ref{hamil}}) that $\ket{I}$ will not only couple to $\ket{B}$, but also to $\ket{D}$. Thus, the closed two-level picture discussed before will no longer be valid. In the second cavity $\ket{B}$ will couple to $\ket{E_+}$ while $\ket{D}$ will couple to $\ket{E_-}$, such that at the end of the process the population will be in a superposition of the states $\ket{E_+}$ and $\ket{E_-}$. Consequently the success probability will be reduced.
On the other hand, a stray electric field will shift the energies of the atomic states $\ket{a}$ and $\ket{b}$ in a similar way, introducing a single-photon detuning. Then even under the two photon resonance condition, the population oscillation between $\ket{I}$ and $\ket{B}$, and between $\ket{B}$ and $\ket{E_{+}}$, will not be complete. Thus again the value of $P$ will be reduced. In Fig. (\ref{fig:EB1}) the success probability is calculated as a function of the deviation from the single and the two-photon resonance condition. 
To quantify the magnetic field intensity, the vacuum Rabi frequency has been taken $g/2\pi=34\rm{MHz}$ \cite{Kim}. Note that the deviation from the two-photon resonance condition affects the success probability  more strongly than a single-photon detuning.

\section{PRACTICAL CONSIDERATIONS}
As a guide for the practical implementation of the model under
investigation we summarize and comment next the main requirements of our
proposal:

\noindent(a) \textit{A $V$-type three-level atom-field interaction}. A
$J=0\longleftrightarrow J=1$ transition could be considered in order to
built-up the $V$-type configuration. Equal Rabi frequencies in each of the
two arms are needed to eventually generate the maximally entangled state
$\ket{E^+}$. Unbalanced Rabi frequencies will produce partially
entangled states. For the optical regime, $^{88} Sr$ \cite{strontium}
could be considered as a possible candidate since it presents an
inter-combination line $5 ^{1}S_0 - 5 ^{3}P_1$ that spans
in two symmetric optical transitions in a $V$-type configuration. Note
that a $\Lambda$-type configuration could be also used, provided one
takes $a^{\dag}_{2+} a^{\dag}_{2-} \ket{\Omega}$ as the initial state of
the process.

\noindent(b) \textit{Preparation of the initial Fock-state for the
cavity modes}.
The first step of the proposal requires to prepare cavity $1$ into a
well defined Fock state with one photon in each polarization mode. Such
step could be achieved in the microwave regime by projecting the
cavity field, after being entangled with an atom, into a Fock state via
ionization measurement of the state of the atom \cite{micromaser}. In
the optical domain, a STIRAP-type procedure by means of a three-level
atom interacting with a strong laser field and the vacuum modes of the
first cavity could be used to prepare the initial Fock state in a similar
manner as in the single photon proposal of A.~Kuhn {\it et al.}~\cite{optfock}.
In our case, however, one should consider the good cavity limit
since the generated photons should remain in the cavity
for the the whole entangling process.

\noindent(c) \textit{Switching on/off the interaction between the atom
and the cavity modes
mode}. Typically, in microwave experiments the interaction time is
adjusted by sending the atom through the cavity at a very well
controlled velocity. Assuming identical cavities, interaction times in
each of them should differ by a factor $t_{2}/t_{1}=\sqrt{2}$ which
means that a mechanism to switch on/off the interaction is needed. One
possible solution could consist in adjusting the atomic velocity to the
largest interaction time, i.e., $t_2$, and to switch on an electric
field producing a large enough atomic Stark shift on cavity $1$ to
control and decrease the interaction time with its cavity modes.
Alternatively,
one could consider equal interaction times and adjust the strength of
the two interactions
to $g_{1}=\sqrt{2}g_{2}$ via the cavity volumes
$g\propto1/\sqrt{V}$.
 
\noindent(d) \textit{The strong coupling limit}.
To obtain high $P$, $F$, and $S$ values, the strong coupling limit given by $g \gg \kappa , \Gamma$ is needed. 
In particular, taking $\kappa=0.053g$ and $\Gamma=0.08g$, which corresponds to the best combination of atomic and cavity decay rates of state-of-the-art optical implementations \cite{bestatomic,sauerlong}, one obtains $P=0.41$, $F=0.91$ and $S=2.69$. On the other hand, in the microwave regime one would obtain ($P$,$\,F$,$\,S$) close to the ideal values ($1$,$\,1$,$\,2\sqrt{2}$) for current experimental parameter values.

\section{CONCLUSIONS}
In conclusion, we have proposed a scheme for the deterministic
generation of polarization entangled photon pairs based on the
interaction of a three-level atom with the two opposite circular polarization
modes of two high-Q cavities. The entangling mechanism consists in the
implementation of two spatially separated $\pi$-Rabi oscillations with
the polarization modes of each cavity. After the interaction with the
cavities, the atomic state decouples from the e.m. field state and the
modes of the two cavities become maximally entangled in their
polarization degree of freedom. By using the MCWF formalism, we have analyzed and characterized this
cavity-QED source in presence of decoherence and experimental
imperfections and discussed some practical considerations for both, the
microwave and optical regimes.

We acknowledge support from the contracts FIS2005-01497MCyT (Spanish Government), SGR2005-00358 (Catalan Government), and EME2004-53 (Universitat Aut\`onoma de Barcelona). KE acknowledges support received from the European Science Foundation PESC QUDEDIS.


\bigskip

\begin{thebibliography}{99}
\bibitem{LHV} A. Aspect, P. Grangier, and G. Roger, "Experimental Realization of Einstein-Podolsky-Rosen-Bohm Gedankenexperiment: A New Violation of Bell's Inequalities," Phys. Rev. Lett. \textbf{49}, 91-94 (1982).
\bibitem{teleportation} C. H.~Bennett, G.~Brassard, C.~Cr\'epeau, R.~Jozsa, A.~Peres, and W. K.~Wootters, "Teleporting an unknown quantum state via dual classical and Einstein-Podolsky-Rosen channels," Phys. Rev. Lett. \textbf{70}, 1895-1899 (1993).
\bibitem{coding} C. H.~Bennett, and S. J.~Wiesner, "Communication via one- and two-particle operators on Einstein-Podolsky-Rosen states," Phys. Rev. Lett. \textbf{69}, 2881-2884 (1992).
\bibitem{QKD} N. Gisin, G. Ribordy, W. Tittel, 
and Hugo Zbinden, "Quantum cryptography," Rev. Mod. Phys. \textbf{74}, 145–195 (2002);
N. L\"utkenhaus, M. Hendrych, and M. Dušek, "Quantum cryptography," to appear in 
Progress in Optics 49, ed. E. Wolf (Elsevier, Amsterdam 2006).

\bibitem{BB84} C. H. Bennet and G. Brassard, Proc. Internat. Conf. Comp. Syst. Signal Proc., Bangalore pp. 175 (1984). 
\bibitem{SARG04}  V. Scarani, A. Acín, G. Ribordy and N. Gisin, "Quantum Cryptography Protocols Robust against Photon Number Splitting Attacks for Weak Laser Pulse Implementations," Phys. Rev. Lett. \textbf{92}, 057901 (2004). 
\bibitem{Ek91} A. K. Ekert, "Quantum cryptography based on Bell's theorem," Phys. Rev. Lett. \textbf{67}, 661-663 (1991). 
\bibitem{Gr00} A.~Beveratos, R.~Brouri, T.~Gacoin, A.~Villing, J-P.~Poizat and P.~Grangier, "Single Photon Quantum Cryptography," Phys. Rev. Lett. \textbf{89}, 187901 (2002).

\bibitem{QCE1} T.~Jennewein, C.~Simon, G.~Weihs, H.~Weinfurter, and A.~Zeilinger, "Quantum Cryptography with Entangled Photons," Phys. Rev. Lett. \textbf{84}, 4729-4732 (2000). 
\bibitem{QCE2} D. S.~Naik, C. G.~Peterson, A. G.~White, A. J.~Berglund, and P. G.~Kwiat, "Entangled State Quantum Cryptography: Eavesdropping on the Ekert Protocol," Phys. Rev. Lett. \textbf{84}, 4733-4736 (2000). 
\bibitem{QCE3} W.~Tittel, J.~Brendel, H.~Zbinden, and N.~Gisin, "Quantum Cryptography Using Entangled Photons in Energy-Time Bell States," Phys. Rev. Lett. \textbf{84}, 4737-4740 (2000). 
\bibitem{ZSS05} D. L.~Zhou, B.~Sun, C. P.~Sun, and L.~You, "Generating entangled photon pairs from a cavity-QED system," Phys. Rev. A {\bf72}, 040302 (2005). 
\bibitem{YYG05} L. Ye, L.-B. Yu, and G.-C. Guo, Phys. Rev. A \textbf{72}, 034304 (2005).
\bibitem{Har} A.~Rauschenbeutel, G.~Nogues, S.~Osnaghi, P.~Bertet, M.~Brune, J. M.~Raimond, and S.~Haroche, "Step-by-step
engineered multiparticle entanglement," Science \textbf{288}, 2024-2028 (2000).
\bibitem{Wal} H.~Walther, "Generation of Photon Number States on Demand," Fortschr. Phys. {\bf 51}, 521-530 (2003).
\bibitem{Rem} G.~Rempe, R.J.~Thompson, H.J.~Kimble, and R.~Lalezari, "Measurement of Ultralow Losses in an Optical Interferometer," Optics Letters {\bf 17}, 363-365 (1992).
\bibitem{Kug} Y. Shimizu, N. Shiokawa, N. Yamamoto, M. Kozuma, T. Kuga, L. Deng, and E. W. Hagley, "Control of Light Pulse Propagation with Only a Few Cold Atoms in a High-Finesse Microcavity," Phys. Rev. Lett. {\bf 89}, 233001 (2002).
\bibitem{Kim} R.~Miller, T.~E.~Northup, K.~M.~Birnbaum, A.~Boca, A.~D.~Boozer and H.~J.~Kimble, 
"Trapped atoms in cavity QED: coupling quantized light and matter," J. Phys. B. {\bf 38} S551-S556 (2005).
\bibitem{Oro} G. T.~Foster, S. L.~Mielke, and L. A.~Orozco, "Intensity correlations in cavity QED," Phys. Rev. A. {\bf 61}, 053821 (2000) .
\bibitem{Hei} T.~Legero, T.~Wilk, M.~Hennrich G.~Rempe, and A.~Kuhn, "Quantum Beat of Two Single Photons, "Phys. Rev. Lett. \textbf{93}, 070503 (2004).
\bibitem{Mor05} G.~Morigi, J.~Eschner, S.~Mancini, and D.~Vitali, "Entangled light pulses from single cold atoms," Phys. Rev. Lett. {\bf96}, 023601 (2006).
\bibitem {optfock} A.~Kuhn, M.~Henrich, and G.~Rempe, "Deterministic Single-Photon Source for Distributed Quantum Networking," Phys. Rev. Lett. {\bf 89}, 067901 (2002).
\bibitem{Ari} E. Arimondo, {\it Coherent population trapping in laser spectroscopy}, Progress in Optics {\bf 35}, 257 (1996).
\bibitem{libro}Y.R.~Shen, "The principles of Non-Linear Optics," (Wiley-Intersecience Publication, 1984).
\bibitem{QJ1} J.~Dalibard, Y.~Castin and K.~M{\o}lmer, "Wave-function approach to dissipative processes in quantum optics," Phys. Rev. Lett. {\bf 68}, 580-583 (1992).
\bibitem{CHSH} J.F.~Clauser, M.A.~Horne, A.~Shimony, and R.A.~Holt, "Proposed Experiment to Test Local Hidden-Variable Theories," Phys. Rev. Lett. {\bf 23}, 880-884 (1969).
\bibitem{sauerlong}J. A.~Sauer, K. M.~Fortier, M. S.~Chang, C. D.~Hamley and M. S.~Chapman, "Cavity QED with optically transported atoms," Phys. Rev. A {\bf 69}, 051804 (2004). 
\bibitem{strontium} 
M.~Takamoto and H.~Katori, "Spectroscopy of the $^{1}$S$_{0}$–$^{3}$P$_{0}$ Clock Transition of $^{87}$Sr in an Optical Lattice," Phys. Rev. Lett. {\bf 91}, 223001 (2003); G.~Ferrari, P.~Cancio, R.~Drullinger, G.~Giusfredi, N.~Poli, M.~Prevedelli, C.~Toninelli, and G. M.~Tino, "Precision Frequency Measurement of Visible Intercombination Lines of Strontium," Phys. Rev. Lett. {\bf 91}, 243002 (2003); H.~Katori, T.~Ido, Y.~Isoya, and M.~Kuwata-Gonokami," Magneto-Optical Trapping and Cooling of Strontium Atoms down to the Photon Recoil Temperature, "Phys. Rev. Lett. {\bf 82}, 116-1119 (1999).
\bibitem{micromaser} B.T.H.~Varcoe, S.~Brattke, and H.~Walther, "Generation of Fock states in the micromaser," J. Opt. B. {\bf 2}, 154-157 (2000).
\bibitem{bestatomic} T. E.~Northup, K.~M.~Birnbaum, A.~Boca, A.D.~Boozer, J.~McKeever, R.~Miller and H.J.~Kimble. In L.G.~Marcassa, V.S.~Bagnato, and K.~Helmerson (editors), {\it Atomic Physics 19}, volume 770, Am. Inst. Phys., New York (2004), pp 313-314. 
\end{thebibliography}
\end{document}